\documentclass[preprint, authoryear, 3p]{elsarticle}

\usepackage{amsmath,amssymb,amsthm}
\usepackage{graphicx}
\usepackage{epstopdf}

\usepackage{hyperref}
\usepackage[pagewise]{lineno}

\def\<{\langle}
\def\>{\rangle}
\def\be{\begin{equation}}
\def\ee{\end{equation}}
\def\bea{\begin{eqnarray}}
\def\eea{\end{eqnarray}}

\begin{document}






\bibliographystyle{model5-names}\biboptions{authoryear}




\title{Self-organization of grid fields under supervision of place cells in a neuron model with associative plasticity}


\author{Andrey Stepanyuk\\
Bogomoletz Institute of Physiology, Kiev, Ukraine, standrey@biph.kiev.ua}

%

\begin{abstract}
The grid cells (GCs) of the medial entorhinal cortex (MEC) and place cells (PCs) of the hippocampus are assumed to be the key elements of the brain network for the metric representation of space. Existing theoretical models of GC network rely on specific hypotheses about the network connectivity patterns. How these patterns could be formed during the network development is not fully understood. It was previously suggested, within the feedforward network models, that activity of PCs could provide the basis for development of GC-like activity patterns. Supporting this hypothesis is the finding that PC activity remains spatially stable after disruption of the GC firing patterns and that the grid fields almost disappear when hippocampal cells are deactivated. Here a new theoretical model of this type is proposed, allowing for grid fields formation due to synaptic plasticity in synapses connecting PCs in hippocampus with neurons in MEC. Learning of the hexagonally symmetric fields in this model occurs due to complex action of several simple biologicaly motivated synaptic plasticity rules. These rules include associative synaptic plasticity rules similar to BCM rule, and homeostatic plasticity rules that constrain synaptic weigths. In contrast to previously described models, a short-term navigational experience in a novel environment is sufficient for the network to learn GC activity patterns. We suggest that learning on the basis of simple and biologically plausible associative synaptic plasticity rules could contribute to the formation of grid fields in early development and to maintenance of normal GCs activity patterns in the familiar environments.
\end{abstract}

%
%

\maketitle

\section{Introduction}

The grid cells (GCs) in the medial entorhinal cortex (MEC) of mammals are the neurons whose firing activity during animal navigation is concentrated near the centers of hexagonally symmetric grid covering the environment \citep{Hafting_2005, Fyhn_2007, Stensola_2012, Moser_annual_rev_neurosci_2008, Moser_2015_ColdSprHarb}. The network of GCs is considered to be a component of the brain system for metric representation of space and navigation \citep{Moser_annual_rev_neurosci_2008, Hartley_2014, Moser_2015_ColdSprHarb}. Patterns of activity, connectivity and development of GC networks have been extensively investigated, and a number of theoretical models of GC network was developed during last decade (see \citep{Zilli_2011} and \citep{Moser_2014_PhTrRSocLond_network_mechanisms} for review and  \citep{Pilly_2013, Castro_2014, Widloski_2014, Bush_2014, Hasselmo_2014} for recent models). These models differ significantly in the way they represent current position of the animal. For example, models based on continuous attractor networks (CAN) \citep{Conklin_2005, McNaughton_2006, OKeefe_2005, Fuhs_2006}, interference of slow oscillations \citep{Blair_2007, Burgess_2007, Gaussier_2007, Giocomo_2007} and feedforward networks \citep{Kropff_Treves_2008, Castro_2014, Pilly_2013} exist. Depending on the nature of input signals these models could be divided into path-integrating models and models that utilize location-specific combinations of input signals. Currently, none of the existing GC network models could be considered as sufficiently complete and supported by the experimental data.

Most of the GC network models could be assigned to the path-integrating type. They imply the emergence of a network, controlled by the signals about animal's speed and direction of movement, in order to provide integration of movements along the path \citep{Zilli_2011, Moser_2014_PhTrRSocLond_network_mechanisms}. Models of this type have a number of restrictions. For example, path integrating models use only a small part of sensory input signals from those which are received by EC and could be the source of navigational information. Other set of problems arise because path integration process in these models is assumed to be continuous, and gradually accumulating integration errors should constantly be adjusted by some additional mechanism \citep{Hardcastle_2015, Cheung_2012}.

Most of the existing models of path-integrating type rely on relatively complex predefined patterns of network connectivity and do not address the problem of these patterns development. For example, many path-integrating models, which use CAN for animal position representation \citep{Conklin_2005, McNaughton_2006, OKeefe_2005, Fuhs_2006}, are based on hypotheses that 1) two-dimentional neuronal network with toroidal topology of synaptic connections exists in the entorhinal cortex, 2) strength and symmetry of these connections are high enough for a stable localized bump or a periodic set of bumps of activity to exist in the network, and 3) additional groups of neurons exist with activity, modulated by speed and direction of animal movement, and with asymmetric connectivity within attractor network layer. As a result, the bump of activity in 2D layer of neurons could move in a coordinated manner with animal movement. Several possible scenarios of development of path integrating CAN were proposed \citep{Stringer_2002, Hahnloser_2003, McNaughton_2006, Widloski_2014}. The problem of development of most of connectivity patterns, assumed by hypotheses 1)-3), was directly addressed in a recent model of Widloski and Fiete \citep{Widloski_2014}. In their work, formation of path-integrating 2D network of GCs under control of both location-specific and velocity-modulated input signals was demonstrated. This network was obtained as a result of self-organization process driven by spike-time dependent plasticity (STDP) in the lateral synaptic connections in the network of excitatory and inhibitory neurons. 

Another type of the GC models suggests that the feedforward network connecting neurons with spatially modulated patterns of activity, such as hippocampal place cells (PCs) or MEC/parasubicular stripe cells, to GCs is formed due to associative synaptic plasticity. \citep{Kropff_Treves_2008,Si_Treves_2012, Pilly_2013, Castro_2014}. 
In several models of this type, formation of grid fields in a given
environment is associated with formation of synaptic connections to GCs selectively from those PCs whose firing fields are located in a nodes of hexagonal lattice \citep{Kropff_Treves_2008, Castro_2014}. This connectivity pattern is supported by the finding that PC activity remains spatially stable after the disruption of GC firing patterns \citep{Koenig_2011, Hales_2014, Bush_Barry_2014} and that grid patterns almost disappear when hippocampal cells are deactivated \citep{Bonnevie_2013}. Development of the early PCs before GCs formation also supports the role of PCs as spatial information providers to GCs.\citep{Langston_2010,Wills_2010}

The limitation of models based on PC-GC feedforward network self-organization \citep{Kropff_Treves_2008, Castro_2014} is that they do not explain explicitly how normal pattern of GC activity could persist during animal's navigation in darkness \citep{Hafting_2005}. Other GC network property which seems to be difficult to explain within PC-GC feedforward network hypothesis is the fact that in a novel environment relative phases of grid fields in the pairs of GCs do not change, while place fields rebuilt completely and their relative phases in PC pairs before and after remapping are not correlated \citep{Fyhn_2007}. As it was described in the work of Kropff and Treves \citep{Kropff_Treves_2008}, alignment of grid fields orientations in the network and correlated remapping of grid fields in different environments could be, in principle, achieved if GCs activity before grid fields formation is modulated by head direction cells and if the strength of horizontal excitatory synaptic connections, which form between pairs of GCs in 2D layer of GCs, is modulated by the degree of similarity of their prefered head directions and by their relative distance in 2D layer. In the work of Si, Kroppf and Treves \citep{Si_Treves_2012} it was shown that coordinated remapping of grid fields between two environments indeed can be observed in the model if the sets of presynaptic PCs active in these environments are weakly overlaping. 
Modular organization of GC networks, and the fact that formation of new grid fields in a novel environment usually do not require prolonged sensory experience in this environment, are the other issues that require explanation within the framework of feedforward network models \citep{Barry_2007, Stensola_2012, Barry_2012}.

Models based on PC-GC feedforward network self-organization rely on specific plasticity rules and differ by their characteristics. For example, the model of Kropff and Treves \citep{Kropff_Treves_2008,Si_Treves_2012} assumes that the mechanism of GCs firing rate adaptation besides Hebbian plasticity rules in PC-GC synapses is important for grid fields development. 
The model of Castro and Aguiar \citep{Castro_2014} assumes the existence of a specific complex rule linking synaptic states and animal's position. 
At the same time, it is well-known that the patterns with hexagonal symmetry are one of the most ubiquitous types of patterns that emerge as a result of symmetry-breaking bifurcations in many physical systems \citep{Stewart}, in particular in the models of self-organization based on Hebbian plasticity in neural networks described by the neural fields equations \citep{Ermentrout_1979, Bressloff_2001}. Thus, it is interesting to explore other possible mechanisms of grid fields development, with a more general types of plasticity rules involved.

The goal of this work is to describe a class of models in which the grid fields are formed as a result of plasticity of PC-GC synapses or synapses from sensory neurons to GC, and to demonstrate a number of possible grid fields formation scenarios in these models.\\
The main difference of the model proposed in this paper from the previously proposed models is that it uses only simplest rules of synaptic plasticity, similar to those supported by experimental data, and does not require assumptions about the firing rate adaptation. 

The most important assumption made in this work is that dependence of associative plasticity rate of PC-GC synapses on the frequency of presynaptic action potentials has a minimum within a typical range of PCs activity during animal's exploration of environment. This assumption is supported by a number of experimental works demonstrating that transition from long term depression (LTD) at low pre- and postsynaptic firing rates to long term potentiation (LTP) at high rates could occur in hippocampal \citep{Bear_1994,Artola_1993, Dudek_1992, Mayford_1995, Wang_1999} and entorhinal cortex synapses \citep{Solger_2004, Zhou_2005, Deng_2007}. The second assumption is that some components of associative or homeostatic synaptic plasticity depend nonlinearly on synaptic weights, and that the curvature of this dependence is positive. This assumption seem to be physiologically relevant since it simply results, for example, from lower-bounding of exitatory synapses weights.          
   
Learning of grid fields in the proposed model could be very fast and does not require a lot of exploratory experience in a novel environment. In addition, the proposed model is compatible with CAN models of GC network and could serve as a basis for the path integration error correction, and for the self-organization of activity bump position control in them.

\section{Theory}

\subsubsection{Description of the model}

In this work we develop a hypothesis that grid cells recognize sets of sensory signals, associated with location of an animal in certain places of environment, with the help of clusters of synapses from hippocampal place cells to grid cell or from neurons of different sensory systems to grid cell. Clusters are composed of synapses with correlated presynaptic activity and form as a result of Hebbian associative plasticity. In addition, under the action of both associative synaptic plasticity and homeostatic plasticity, which limits weights of the individual synapses or groups of synapses, separation of clusters occurs in such a way that different clusters encode maximally different sets of local sensorу signals.

As a possible mathematical model of such a system we consider a network of neurons with linear activation rule 
\begin{eqnarray} \label{activation}
y_t=\textbf{w}_t^T\textbf{x}_{t}
\end{eqnarray}
where $\textbf{x}_t $ is a vector of firing rates of presynaptic neurons at time moment $t $, $\textbf{w}_t $ – vector of PC-GC synaptic weights and $y_t $ - firing rate of postsynaptic neuron.

We will consider several types of synaptic plasticity rules which will lead to different scenarios of grid fields self-organization.
Let us start from the general form of equation describing synaptic weights change as a function of rates of presynaptic and postsynaptic action potentials: 
\begin{eqnarray} \label{learning_gen}
 \dot{w}_{it}= \eta\left( x_{it},y_{t}\right) + \zeta\left( y_t,\textbf{w}_t \right)w_{it}+P(w_{it},x_{it})
\end{eqnarray}
where $ \zeta $ and $ P $ terms describe global homeostatic plasticity (restricting sum of weigths of all synapses) and local homeostatic plasticity (restricting weights of individual synapses independently of each other). The $\eta $ term describes associative plasticity corresponding to LTP and LTD.
For simplicity, we approximate homeostatic plasticity terms by polynomials:
\begin{align}  
 \zeta(y_t)=-\sigma y_{t}^2 \quad \text{  or  }  \quad  \zeta(\textbf{w}_t)=-\sigma \sum_{k}{w_{kt}^2}  ,\label{hom_plasticity:1}\\ P(w_{it},x_{it})=A(x_{it})w_{it}+B(x_{it})w_{it}^2 ,\label{hom_plasticity:2}
 \end{align} 
where $ A(0)=0 $, $ B(0)=0 $ and $ \sigma $ is some positive number. Here, we assume that local homeostatic plasticity rate is relatively low, so that third and higher order terms of its Taylor series expansion could be neglected.

From experiments it is known that transition from LTD at low pre- and postsynaptic firing rates to LTP at high rates could occur in hippocampal \citep{Bear_1994,Artola_1993, Dudek_1992, Mayford_1995, Wang_1999} and entorhinal cortex synapses \citep{Solger_2004, Zhou_2005, Deng_2007}. To describe this phenomenon, polynomial approximation of $\eta\left( x_{it},y_{t}\right)  $ should contain terms up to at least third order:
\begin{eqnarray} \label{asslearning_gen}
\eta\left( x_{it},y_{t}\right)=-\eta_- x_{it}y_{t}+ \eta_{2+}x_{it}y_{t}^2 + \eta_+ x_{it}^2y_{t}  
\end{eqnarray}
where coefficients $ \eta_- $ (rate of LTD), $ \eta_+ $ (rate of LTP linear component) and $ \eta_{2+} $ (nonliner LTP rate) are positive.
Without last term this equation is similar to the Bienenstock-Cooper-Munro (BCM ) rule \citep{pmid7054394,pmid23080416}, phenomenological synaptic plasticity rule predicted theoreticaly to account for selectivity of primary sensory cortex neurons. Similar plasticity rules were obtained by averaging of plastic changes that occur in response to a Poisson stream of pre- and postsynaptic spikes in detailed phenomenological models of STDP \citep{pmid22080608,pmid21423519,pmid20438333}. Other possible interpretations of Eq.(\ref{asslearning_gen}) is that its first and second terms are equivalent to linear and quadratic terms of the Taylor series expansion of anti-Hebbian plasticity rate dependence on postsynaptic activity. This could occur, for example, if neuron has nonlinear activation function $S(z)  $
\begin{eqnarray} \label{nonlin_activ}
y_t=S(z_t)=S(\textbf{w}_t^Tx_{t}) \\ 
\eta\left( x_{it},z_{t}\right)=-\eta_0 x_{it}y_{t}=-\eta_- x_{it}z_{t}+\eta_{2+}x_{it}z_{t}^2 +o(z_{t}^3)   
\end{eqnarray} 
Here $ y_t $ is the output firing rate and $ z_t $  is the average change in postsynaptic membrane potential evoked by synaptic activity. \\
Last term in Eq.(\ref{asslearning_gen}) could be a result of nonlinearity of LTD rate dependence on presynaptic activity. As we will see later, relatively small nonlinearity of this dependence could be sufficient for grid fields formation. There are experimental results which support assumption that this term indeed could be required for precise description of STDP at least at low rates of postsynaptic activity. For example, it was shown that spike time dependent LTP could be observed in absence of postsynaptic spiking \citep{pmid9852584, pmid17065442}. Probably in this case, at low postsynaptic activity levels, the rate of synaptic changes could approximately linearly depend on $ y_{t} $ and nonlinearly on $ x_{it} $. Other motivation for introducing terms with differences in dependence of LTP component of plasticity on presynaptic activity with respect to LTD is experimental results suggesting that LTP and LTD could be selectively supported by different sets of postsynaptic receptors and voltage-gated channels \citep{pmid17065442, Solger_2004}. But, as we will see later, LTP component is not required for development of grid fields.  

\subsubsection{Analysis of weakly nonlinear version of the single neuron model}

To analyze potential mechanism of grid fields formation in single neuron model with Hebbian and anti-Hebbian synaptic plasticity described by Eqs.(\ref{learning_gen})-(\ref{asslearning_gen}) it is convenient to consider a variant of this model which is weakly nonlinear with respect to synaptic weights. 
Namely, we will assume that in Eqs.(\ref{learning_gen})-(\ref{asslearning_gen}) all the terms that are nonlinear with respect to $ \textbf{w}_t $  are small, except for those which are equal for all synapses ( $\sigma(y_t,\textbf{w}_t) \textbf{w}_t $ ).\\
If learning rate is slow with respect to the exploration speed, learning equations Eqs.(\ref{learning_gen})-(\ref{asslearning_gen}) could be integrated over $t$ :
\begin{eqnarray} \label{learning_integral}
  \dot{w}_{i} =  -\eta_- \textbf{R}_{1i}\textbf{w}+\eta_+   \textbf{R}_{2i}\textbf{w} + \eta_{2+}\textbf{w}^T\textbf{Q}_i\textbf{w} + \sigma\left( \textbf{w}\right)w_{i}+P(w_{i})
\end{eqnarray}
Here $\textbf{R}_1$,$\textbf{R}_2$ and $\textbf{Q}_i$ are time averages of  functions of input signals:
 \begin{align} \textbf{R}_1=\frac{1}{T}\sum_{t}{\textbf{x}_t\textbf{x}_t^T}  ,\label{R_1}\\  \textbf{R}_2=\frac{1}{T}\sum_{t}{\textbf{x}_t^2\textbf{x}_t^T} ,\label{R_2}\\ \quad \quad \textbf{Q}_i=\frac{1}{T}\sum_{t}{\textbf{x}_t\textbf{x}_t^Tx_i}, \label{Q_i}\end{align} 
 If the animal explores environment for a long time, and visits all places with equal probability, or if learning rates are normalized according to the extent of place familiarity, then the matrix $\textbf{R}_1$ could be expressed as a function of presynaptic PCs firing fields $F_i(\textbf{r})$ :
\begin{eqnarray} \label{fields_corr_R}
 R_{1ij}\approx \int F_i(\textbf{r})F_j(\textbf{r})dr^D
\end{eqnarray}         
We assume that all place fields are circulary symmetric within some distance from their centers and are approximately equal: 
\begin{eqnarray}  \label{place_fields_f}
 F_i(\textbf{r})\approx f\left( \left\| \textbf{r}-\textbf{r}_i\right\|\right)   
 \end{eqnarray} 
For approximately Gaussian place fields $  F_i(\textbf{r})\approx  F\exp(-\frac{1}{2\sigma_r^2}\left\|\textbf{r}-\textbf{r}_i\right\|^2  ) $ , cross-correlation $\textbf{R}_1$ has simple analytical expression:
\begin{eqnarray} \label{gauss_fields_R}
R_{1ij}\approx 2\pi \frac{\sigma_r^2}{4} F^2\exp(-\frac{1}{4\sigma_r^2}\left\| \textbf{r}_i-\textbf{r}_j \right\|^2) 
\end{eqnarray}
Analogously for $\textbf{R}_2$ and $\textbf{Q} $ :
\begin{eqnarray} \label{gauss_fields_R2_Q}
{R_{2ij}\approx \int F_i(\textbf{r})^2F_j(\textbf{r})dr^D = 2\pi \frac{\sigma_r^2}{6} F^3\exp\left( -\frac{1}{3\sigma_r^2}\left\| \textbf{r}_i-\textbf{r}_j \right\|^2\right)} \\
Q_{ijk}\approx \int F_i(\textbf{r})F_j(\textbf{r})F_k(\textbf{r})dr^D =
\left(R_{1ij}R_{1ik}R_{1jk}\right)^{\frac{2}{3}}
\end{eqnarray}

It is known that development of periodic patterns with different symmetries may be observed in systems described by equations similar to Eq.(\ref{learning_integral}) as a result of symmetry-breaking bifurcations. The symmetry of periodic pattern, which begins to grow in this and more general systems after the bifurcation, can be predicted using neural fields approximation and methods of weakly nonlinear analysis \citep{Bressloff_2005, Ermentrout_1991, pmid23345648}. The stability of stationary periodic patterns with respect to their periodic perturbations could be investigated using methods of symmetry-breaking bifurcation analysis for neural fields equations \citep{Stewart, Bressloff_2001}.

Here we only briefly discuss, using standard approach, the mechanism of formation of spatialy periodic sensory fields with hexagonal symmetry in the system Eq.(\ref{learning_integral}). 
First, note that Lyapunov function exists for Eq.(\ref{learning_integral}):
\begin{eqnarray} \label{energy}
E= \eta_- \frac{1}{2}\textbf{w}^T\textbf{R}_{1}\textbf{w}-\frac{1}{2}\eta_+   \textbf{w}^T\textbf{R}_{2}\textbf{w} - \frac{1}{3}\eta_{2+}\int{\left(\textbf{w}^T\textbf{x}_{\textbf{r}} \right)^3dr^D } + \frac{1}{4}\sigma\left( \textbf{w}^T\textbf{w}\right)^2 -\sum_{i}{\int{P(w_{i})dw_{i}}}
 \end{eqnarray}
As a consequence, all solutions of Eq.(\ref{learning_integral}) should converge to a steady-state solution that provides a local minimum of the Lyapunov function Eq.(\ref{energy}).

This local minimum always exists since the length of weight vector $ L_w=\sqrt{\textbf{w}^T\textbf{w} } $ will decrease for all weigth vectors with the length larger then some limit. For example, for Eq.(\ref{learning_integral}) without third term (i.e. in the case $ \eta_{2+}=0 $ ) this follows from the set of inequalities:
\begin{eqnarray} \label{w_length}
\frac{d}{dt}\left(\frac{1}{2}L_w^2\right) =\textbf{w}^T\dot{\textbf{w}}=\textbf{w}^T\textbf{R}_0\textbf{w}-\sigma\left(\textbf{w}^T\textbf{w} \right)^2+\textbf{w}^TP(\textbf{w}) \le \lambda_{max}L_w^2-\sigma L_w^4+BNw_{max}^3  \nonumber\\ \le \lambda_{max}L_w^2-\sigma L_w^4+BNL_w^3
\end{eqnarray}
where $ \lambda_{max} $ is the maximal eigenvalue of the matrix $\textbf{R}_0= -\eta_-\textbf{R}_1 +\eta_+\textbf{R}_2 +\varepsilon_1\textbf{E} $, $ N $- the numbet of elements in the vector $ \textbf{w} $ and  $w_{max} $ - the element of $ \textbf{w} $ with a maximal absolute value. It can be seen that $ \dot{L_w}<0 $ for all $ L_w $ larger than the largest root of the polinomial in the right side of Eq.(\ref{w_length}). 

Futher analysis will show that the solution of Eq.(\ref{learning_integral}) will always approach grid-like pattern at least for some sets of parameters.
We assume for simplicity that the domain for place fields spatial locations is a twisted torus. In other words, each place field $F(\textbf{r}) $ has periodic structure
\begin{eqnarray} \label{latice_fields_F}
F(\textbf{r})=F(\textbf{r}+n_1\textbf{v}_1L+n_2\textbf{v}_2L)
\end{eqnarray}  
where $n_1, n_2\in\mathbb{Z}$, $\textbf{v}_1=(1,0) $ and $\textbf{v}_2=(\frac{1}{2},\frac{\sqrt{3}}{2}) $ are basis vectors of the hexagonal lattice of size $L$. 
In this case, first three terms in equation Eq.(\ref{learning_integral}) are convolutions. As follows from the convolution theorem, the equation for the spatial Fourier decomposition of weights distribution $ \textbf{w} $ as a function of place fields positions $ \textbf{r} $ has simple form:
\begin{eqnarray} 
  \textbf{w}_{\textbf{r}}=\sum_{\textbf{k}\in\tilde{\Lambda}}^{}{\alpha_{\textbf{k}}e^{i\textbf{k}\textbf{r}}} \label{waves_of_weigths}\\ 
  \dot{\alpha}_{\textbf{k}} =  - \eta_-\tilde{\textbf{R}}_{1\left| \textbf{k} \right|}\alpha_{\textbf{k}} + \eta_+\tilde{\textbf{R}}_{2\left| \textbf{k}\right|}\alpha_{\textbf{k}}  +  \eta_{2+}\sum_{\textbf{k}'}^{}{ \textbf{Q}_{1\left| \textbf{k}\right|,\left| \textbf{k}'\right|,\left| \textbf{k-k}'\right|} \alpha_{\textbf{k}'}\alpha_{\textbf{k-k}'} } \nonumber \\                              \label{waves_evolution_eq}
  - \sigma\sum_{\textbf{k}'}{\left|\alpha_{\textbf{k}'} \right|^2 } \alpha_{\textbf{k}} + P_1\left( {\alpha,\textbf{k}}\right) \\
 P_1\left( {\alpha,\textbf{k}}\right)= \varepsilon_1\alpha_{\textbf{k}}+\varepsilon_2\sum_{\textbf{k}'}^{}{ \alpha_{\textbf{k}'}\alpha_{\textbf{k-k}'} } 
\end{eqnarray}
where wavevector $ \textbf{k} $ is defined on the hexagonal lattice $ \tilde{\Lambda} $ which is dual to the lattice $ \Lambda $, $ \textbf{Q}_{1\left| \textbf{k}\right|,\left| \textbf{k}'\right|,\left| \textbf{k-k}'\right|} =F_q exp\left( -\frac{\sigma_r^2}{2}\left| \textbf{k}\right|^2 -\frac{\sigma_r^2}{2}\left| \textbf{k}'\right|^2 - \frac{\sigma_r^2}{2}\left| \textbf{k+k}'\right|^2\right)  $ for place fields described by Eq.(\ref{gauss_fields_R}), $ \epsilon_1\approx \int A\left(  F(\textbf{r})\right)dr^D $,  $ \epsilon_2\approx \int B\left(  F(\textbf{r})\right)dr^D $.

Let us initially consider evolution of wave amplitude, $ \alpha_{\textbf{k}}  $ , starting from small positive initial values of synaptic weights. If $ \varepsilon_1 $ is positive, then there is some $  \textbf{k} $ for which $ -\eta_-\tilde{\textbf{R}}_{1\left| \textbf{k} \right|}\alpha_{\textbf{k}} + \eta_+\tilde{\textbf{R}}_{2\left| \textbf{k}\right|}\alpha_{\textbf{k}}+\varepsilon_1\alpha_{\textbf{k}}  $ is  positive as well. Since for small $ \alpha_{\textbf{k}}  $ influence of all high order terms in Eq.(\ref{waves_evolution_eq}) can be neglected, then the most rapidly growing wave amplitudes are those located on the circle  $ \left| \textbf{k}\right|=k_c $, where $ k_c $ is the amplitude of wavevector for which expression $ -\eta_-\tilde{\textbf{R}}_{1\left| \textbf{k} \right|} + \eta_+\tilde{\textbf{R}}_{2\left| \textbf{k}\right|}  $ is maximal. As a result of exponential growth of wave amplitudes, to the end of this 'linear' phase only wave amplitudes with $ \left| \textbf{k}\right|=k_c $ will be significantly different from zero, but amplitudes of all these waves grow with approximately equal speed. In other words, the ratios of amplitudes of waves with $ \left| \textbf{k}\right|=k_c $ do not change with time if $ \eta_{2+}=0 $ and $ \varepsilon_2=0 $. 

The influence of second order terms gradually increase with growth of $ \alpha_{\textbf{k}} $, but if $ \eta_{2+} $ and $ \varepsilon_2 $ are small, then most of the wave power will still be concentrated within the narrow ring $\left|  \tilde{\textbf{R}}_{0,\left| \textbf{k}\right|}-\tilde{\textbf{R}}_{0,k_C}\right|\lesssim \varepsilon_2 $ , where $ \tilde{\textbf{R}}_{0,\left| \textbf{k}\right|}= -\eta_-\tilde{\textbf{R}}_{1\left| \textbf{k} \right|} + \eta_+\tilde{\textbf{R}}_{2\left| \textbf{k}\right|} +\varepsilon_1$. 

The action of second order terms on the wave growth generally results in gradual increase of differences in speed of growth of different wave groups, so that only small-sized group of waves will dominate at the end. Indeed, let us consider a group of waves with wavevectors lying on the circle $ \left| \textbf{k}\right|=k_c $. For the first-order approximation, the contribution of all $ \left| \textbf{k}\right|\neq k_c $ to the second order (with respect to $ \alpha_{\textbf{k}} $ ) terms of the equation Eq.(\ref{waves_evolution_eq}) can be neglected. 

For each wavevector $ \textbf{k} $ lying on the circle $ \left| \textbf{k}\right|=k_c $ there is only one triple of wavevectors $ \textbf{k} $, $ \textbf{k}' $, $ \textbf{k-k}' $ with $ \left| \textbf{k}\right|=\left| \textbf{k}'\right|=\left| \textbf{k-k}'\right|=k_c $. Consequently, the equation Eq.(\ref{waves_evolution_eq}) could be simplified: 
\begin{eqnarray}  \label{waves_evol_simplif}
  \dot{\alpha}_{\textbf{k}} =  \left( \tilde{\textbf{R}}_{0,\left| \textbf{k} \right|} -\sigma\sum_{\textbf{k}'}{\left|\alpha_{\textbf{k}'} \right|^2 }\right) \alpha_{\textbf{k}}  +  2 \textbf{Q}_{2,\left| \textbf{k}\right|,\left| \textbf{k}+\right|,\left| \textbf{k}-\right|} \alpha_{\textbf{k}+}\alpha_{\textbf{k}-}     \\
 \textbf{Q}_{2,\left| \textbf{k}\right|,\left|\textbf{k}'\right|,\left| \textbf{k-k}'\right|} =\varepsilon_2\ + \eta_{2+}\textbf{Q}_{1,\left| \textbf{k}\right|,\left| \textbf{k}'\right|,\left| \textbf{k-k}'\right|} 
\end{eqnarray}      
where $ \textbf{k}+  $ and $ \textbf{k}-  $ are wavevectors obtained by rotation of wavevector $\textbf{k} $ by an angle $ \frac{\pi}{3} $ and $ -\frac{\pi}{3} $ , respectively.
Since synaptic weights are real-valued, then amplitude of wave with wavevector $ \textbf{k} $ is equal to the complex conjugate of the amplitude of wave with opposite wavevector: $ \alpha_{-\textbf{k}}=\hat\alpha_{\textbf{k}} $.  
We can observe that differences between absolute values of amplitudes of waves, having wavevectors rotated by the angle $ \frac{\pi}{3} $ with respect to each other, will decrease with time. Indeed, 
let $ \alpha_{\textbf{k}}=|\alpha_{-\textbf{k}}|e^{i\phi_\textbf{k}} $ and let the vectors $ \textbf{k}2+  $ and $ \textbf{k}2-  $ correspond to wavevector $\textbf{k}$ rotated by angles $ \frac{2\pi}{3} $ and $ -\frac{2\pi}{3} $ , respectively. Then from Eq.(\ref{waves_evol_simplif}) it follows that:
\begin{align} 
\frac{d}{dt}|\alpha_{\textbf{k}}|=  \left( \tilde{\textbf{R}}_{0,\left| \textbf{k} \right|} -\sigma\sum_{\textbf{k}'}{\left|\alpha_{\textbf{k}'} \right|^2 }\right)\left| \alpha_{\textbf{k}}\right|  +  2 \textbf{Q}_{2,\left| \textbf{k}\right|,\left| \textbf{k}2+\right|,\left| \textbf{k}2-\right|} \left|\alpha_{\textbf{k}2+}\right|\left|\alpha_{\textbf{k}2-}\right|Re\left\lbrace e^{-i\left( \phi_{\textbf{k}} + \phi_{\textbf{k}2+} + \phi_{\textbf{k}2-} \right) }  \right\rbrace  , \label{abs_alpha:1}\\
\dot{\phi}_{\textbf{k}}  = - 2 \textbf{Q}_{2,\left| \textbf{k}\right|,\left| \textbf{k}2+\right|,\left| \textbf{k}2-\right|}   \left|\alpha_{\textbf{k}}\right|^{-1}\left|\alpha_{\textbf{k}2+}\right|\left|\alpha_{\textbf{k}2-}\right|Im\left\lbrace e^{i(\phi_{\textbf{k}}+\phi_{\textbf{k}2+} +\phi_{\textbf{k}2-})}  \right\rbrace , \label{abs_alpha:2}
\end{align}
As a result, at $ t\longrightarrow\infty $ the sum of phases $ \phi_{\textbf{k}}+\phi_{\textbf{k}2+} +\phi_{\textbf{k}2-}\longrightarrow 0 $ and Eqs.(\ref{abs_alpha:1}) converge to:
\begin{align} 
\frac{d}{dt}|\alpha_{\textbf{k}}|=  \left( \tilde{\textbf{R}}_{0,\left| \textbf{k} \right|} -\sigma\sum_{\textbf{k}'}{\left|\alpha_{\textbf{k}'} \right|^2 }\right)\left| \alpha_{\textbf{k}}\right|  +  2 \textbf{Q}_{2,\left| \textbf{k}\right|,\left| \textbf{k}2+\right|,\left| \textbf{k}2-\right|} \left|\alpha_{\textbf{k}2+}\right|\left|\alpha_{\textbf{k}2-}\right|  , \label{abs_alpha:3}
\end{align}
This equation describes growth of absolute values of amplitudes, that will continue at least while $  \tilde{\textbf{R}}_{0,\left| \textbf{k} \right|} -\sigma\sum_{\textbf{k}'}{\left|\alpha_{\textbf{k}'} \right|^2 }>0 $. From some time moment $ T_d $ the last inequality will be broken for all $ t>T_d $ and differences between absolute values of wave amplitudes for triplets of wavevectors $ (\textbf{k}2 , \textbf{k}2- , \textbf{k}2+) $ will decrease. Indeed, from Eq.(\ref{abs_alpha:3}) we have:
\begin{align} 
\frac{d}{dt}\left( |\alpha_{\textbf{k}}|-|\alpha_{\textbf{k}2+}| \right) = \left(  \tilde{\textbf{R}}_{0,\left| \textbf{k} \right|} -\sigma\sum_{\textbf{k}'}{\left|\alpha_{\textbf{k}'} \right|^2 } -  2 \textbf{Q}_{2,\left| \textbf{k}\right|,\left| \textbf{k}2+\right|,\left| \textbf{k}2-\right|} \left|\alpha_{\textbf{k}2-}\right| \right) \left( \left|\alpha_{\textbf{k}}\right|- \left|\alpha_{\textbf{k}2+}\right|\right)  , \label{abs_alpha:4}
\end{align}
Since expression in left brackets on the right side of Eq.(\ref{abs_alpha:4}) is always negative for $ t>Td $ then $ |\alpha_{\textbf{k}}|-|\alpha_{\textbf{k}2+}|\longrightarrow 0 $ with time. As a result, in a steady-state all nonzero wave amplitudes should be equal.


In general case, the number of wavevectors defined on the hexagonal lattice $ \tilde{\Lambda} $ and lying on the circle $ \left\| \textbf{k} \right\| = k_c $, is an arbitrary natural number divisible by six. But only the configuration with six nonzero waves whose wavevectors compose equilateral hexagon is stable. 

To show this, let us consider evolution of ratios of wave amplitudes $ \mu_{\textbf{k}}=\frac{\alpha_{\textbf{k}}}{\alpha_{\textbf{k1}}} $, where $ \alpha_{\textbf{k1}} $ is the amplitude of the wave having maximal amplitude at $ t\rightarrow\infty $  :
\begin{eqnarray} \label{waves_ratio_evol_simplif}
  \dot{\mu}_{\textbf{k}}  =  \alpha_{\textbf{k1}}\left(   - 2 \textbf{Q}_{2,k_C,k_C,k_C} \mu_{\textbf{k1}+}\mu_{\textbf{k1}-}  \mu_{\textbf{k}}  +  2 \textbf{Q}_{2,k_C,k_C,k_C} \mu_{\textbf{k}+}\mu_{\textbf{k}-}  \right)
    \end{eqnarray} 
  As follows from Eqs.(\ref{abs_alpha:1} - \ref{abs_alpha:4}), at $ t\rightarrow\infty $ this expression approaches: 
  \begin{eqnarray}
  \dot{\mu}_{\textbf{k}}  =  2 \textbf{Q}_{2,k_C,k_C,k_C} \alpha_{\textbf{k1}}\mu_{\textbf{k}}\left( \mu_{\textbf{k}} - 1\right)   
 \end{eqnarray}   
This equation shows that if at some moment of time $ \left| \mu_{\textbf{k}} \right|  $ becomes less than 1, for example due to some small perturbation, then if $\textbf{k}  $ do not belong to the group of wavevectors obtained by rotations of $ \textbf{k1} $ by angles $\frac{\pi}{3}n  $, $ n\in\mathbb{Z} $ then $ \dot{\left|\mu_{\textbf{k}} \right|}  $ will be negative until $ \alpha_{\textbf{k}}  $ has reached zero.

We can conclude that for sufficiently small $ \eta_{2+} $ and $ \varepsilon_2 $ the solution of Eq.(\ref{waves_evolution_eq}) , which starts from small positive weights, will approach a pattern that is close to the hexagonally symmetric sum of three plane waves:\\ 
\begin{eqnarray} \label{solution}
 w_{\textbf{r}}=\alpha Re\left\lbrace \left(e^{i(\textbf{r}-\textbf{r}_0)\textbf{k}} +e^{i(\textbf{r}-\textbf{r}_0)\textbf{k}2-}+e^{i(\textbf{r}-\textbf{r}_0)\textbf{k}2+}\right)  \right\rbrace   
\end{eqnarray}

The same proposition is true for any other starting distribution of synaptic weights. Using first-order perturbation approximation with respect to $ \eta_{2+} $ and $ \varepsilon_2 $ it could be shown that steady-state solution of Eq.(\ref{waves_evolution_eq}) can be approximated by steady-state solution of Eq.(\ref{waves_evol_simplif}) and that hexagonally symmetric sum of plane waves gives the only stable solution of Eq.(\ref{waves_evol_simplif}) and Eq.(\ref{waves_evolution_eq}) for sufficiently small $ \eta_{2+} $ and $ \varepsilon_2 $ 

If the size of environment is large with respect to the size of place fields, and coefficients near second order terms are relatively large, then deviations from the twisted torus assumption will not significantly disturb general form of solution of equation Eq.(\ref{waves_evolution_eq}), since in this case triples of wavectors lying sufficiently close to $ \left| \textbf{k}\right|=\left| \textbf{k}'\right|=\left| \textbf{k-k}'\right|=k_c $ always could be found for each wavector $ \textbf{k} $ lying within the ring of large amplitude waves. As a result, Eqs.(\ref{waves_evol_simplif} - \ref{solution}) will be approximately correct if $  \textbf{Q}_{2,\left| \textbf{k}\right|,\left|\textbf{k}'\right|,\left| \textbf{k-k}'\right|}  $ is sufficiently large.  

The size of grid fields formed in the system described by Eq.(\ref{learning_integral}) is determined by the critical wavelength $ l_c=2\pi/k_c $. It has a simple physiological meaning: it is equal to a twice distance from the center of place field to the position of the animal where rate of associative synaptic plasticity is minimal. For example, in the case when $ \eta_2+=0 $, $ l_c $ is defined by the distance to a place where the presynaptic firing rate is equal to the position of the minimum of associative synaptic plasticity rate dependence on presynaptic firing frequency: $ l_c=2argmin_{\left\| \textbf{r} \right\|}(-\eta_- R_1(\left\| \textbf{r} \right\|) +\eta_+ R_2(\left\| \textbf{r} \right\|)) $. In this case grid size depends only on the size and shape of place fields and on $ \eta_-/\eta_+ $ ratio.   

\subsubsection{Alignment of grid field orientations in models of GCs network}
 
A network of GCs interconnected by a stable synaptic connections with 2D topology and described by Eqs.(\ref{activation} -\ref{asslearning_gen})  could be modeled using activation equations:
 \begin{eqnarray} \label{act_net}
 \dot{y}_{gt}=-\frac{1}{\tau}y_{gt} +\textbf{J}_{g}\textbf{y}_t+\textbf{w}_{gt}^T\textbf{x}_{t}
 \end{eqnarray}
where $ \textbf{J}_{gg1}=J(\left\| \textbf{s}_g-\textbf{s}_{g1} \right\| ) $ is connectivity matrix of the network and $ \textbf{s}_g $ is a position of neuron $ g $ in the 2D layer of GCs, $ \tau $ is membrane time constant of GC, which is small with respect to the time constants of learning. If activity of PCs changes slowly with respect to $ \tau $, and the strength of excitatory synapses is not too high, then $ \dot{y}_{gt}\approx 0 $ and:
 \begin{eqnarray} \label{act_net:2}
 \textbf{y}_{t}\approx\left( \frac{1}{\tau}\textbf{E} -\textbf{J}\right)^{-1}\textbf{w}\textbf{x}_{t}
 \end{eqnarray}
where $ \textbf{w} $ is the connectivity matrix for PC-GC synapses. This approximation assumes that all eigenvalues of the matrix $ \textbf{C}=\left( \frac{1}{\tau}\textbf{E} -\textbf{J}\right)^{-1}  $ are negative.
The learning equations for GC network could be obtained from
Eqs.(\ref{activation})-(\ref{asslearning_gen}). After setting  $  \zeta(\textbf{w}_t)=-\sigma \sum_{g,k}{w_{gkt}^2}$ and  $ P(w_{git},x_{it})=BH(x_{it}-\theta)w_{git}^2  $ in Eqs.(\ref{hom_plasticity:1})-(\ref{hom_plasticity:2}) we obtain:
\begin{eqnarray} \label{learning_net}
 \dot{w}_{git}= -\eta_- x_{it}y_{gt}+ \eta_{2+}x_{git}y_{gt}^2 + \eta_+ x_{git}^2y_{gt}   -\sigma \sum_{g',k}{w_{g'kt}^2}w_{git}+Bw_{git}^2H\left(x_{git}-\varTheta \right) 
\end{eqnarray}
where $ H $ is the Heaviside function. 
Integration of this equation over time gives, similarly to Eq.(\ref{learning_integral}):
\begin{eqnarray} \label{learning_int_net}
  \dot{\textbf{w}_i} =  -\eta_- \textbf{C}\textbf{w}\textbf{R}_{1i}+\eta_+   \textbf{C}\textbf{w}\textbf{R}_{2i} + \eta_{2+}Tr\left\lbrace \textbf{C}\textbf{w}\textbf{Q}_i\textbf{w}^T\textbf{C}^T\right\rbrace   + \sigma\left( \textbf{w}\right)\textbf{w}_i+P(\textbf{w}_i)
\end{eqnarray} 
For the simplicity of futher analysis let us consider $ \eta_{2+}=0 $ case. Using the same assumptions as for derivation of Eqs.(\ref{waves_evolution_eq}) after Fourier decomposition of Eq.(\ref{learning_int_net}) we obtain:
\begin{eqnarray} 
  w_{g\textbf{r}}=\sum_{\textbf{k}\in\tilde{\Lambda}}^{}{\beta_{g\textbf{k}}e^{i\textbf{k}\textbf{r}}} \\ \label{waves_weights_net}
  \dot{\beta}_{g\textbf{k}} =  \textbf{C}\tilde{\textbf{R}}_{0\left| \textbf{k} \right|}\beta_{\textbf{k}}  +  \epsilon_2\sum_{\textbf{k}'}^{}{ \beta_{g\textbf{k}'}\beta_{g\textbf{k-k}'}}                 
  - \sigma\sum_{g',\textbf{k}'}{\left|\beta_{g'\textbf{k}'} \right|^2 } \beta_{g\textbf{k}}  \label{waves_evolution_net}
\end{eqnarray}
If we assume that similarly to the layer of PCs, the GCs positions in 2D layer compose a hexagonal lattice $ \Lambda_s $, and the domain for the Eq.(\ref{learning_int_net}) in the GCs layer is a twisted torus, then for the Fourier decomposition of function $ \beta_{g\textbf{k}} $ over the dual lattice $ \tilde{\Lambda_s} $ we have:
\begin{eqnarray} 
  \beta_{\textbf{s}\textbf{k}}=\sum_{\textbf{m}\in\tilde{\Lambda_s}}^{}{\alpha_{\textbf{m}\textbf{k}}e^{i\textbf{m}\textbf{s}}} \\ \label{waves_net:2}
  \dot{\alpha}_{\textbf{m}\textbf{k}} =  \tilde{\textbf{C}}_{\left| \textbf{m} \right|}\tilde{\textbf{R}}_{0\left| \textbf{k} \right|}\alpha_{\textbf{m}\textbf{k}}  +  \epsilon_2\sum_{\textbf{k}'}^{}{ \alpha_{\textbf{m}'\textbf{k}'}\alpha_{\textbf{m-m}'\textbf{k-k}'}}                 
  - \sigma\sum_{\textbf{m}',\textbf{k}'}{\left|\alpha_{\textbf{m}'\textbf{k}'} \right|^2 } \alpha_{\textbf{m}\textbf{k}}  \label{waves_evolution_net:2}
\end{eqnarray}
The simplest case for the analysis of the pattern formation in system Eq(\ref{waves_evolution_net:2}) is the case when connectivity matrix $ \textbf{J} $  has a maximum at the main diagonal, or in other words function $ J(\left\| \textbf{s}_g-\textbf{s}_{g1} \right\|) $ has maximum at $ \left\| \textbf{s}_g-\textbf{s}_{g1} \right\|=0 $. Note that this condition imposes no restriction on the sign of synaptic weights and thus such connectivity pattern is possible for both purely excitatory and inhibitory GC-GC synaptic connections. 
In the case of this connectivity pattern, the function $ \tilde{\textbf{C}}_{\left| \textbf{m} \right|}  $ will reach a maximum at $ \textbf{m}=0 $ , and as a result, the most rapidly growing wave amplitudes are $ \alpha_{0\textbf{k}} $ for the wavevector $ \textbf{k} $ located on the circle  $ \left| \textbf{k}\right|=k_c $. Using arguments similar to those used for derivation of Eqs.(\ref{waves_evol_simplif} - \ref{solution}) we could obtain a set of similar equations:
\begin{eqnarray}  \label{waves_evol_simplif_net:1}
  \dot{\alpha}_{0\textbf{k}} =  \left(\tilde{\textbf{C}}_0 \tilde{\textbf{R}}_{0,\left| \textbf{k} \right|} -\sigma\sum_{\textbf{m}',\textbf{k}'}{\left|\alpha_{\textbf{m}'\textbf{k}'} \right|^2 }\right) \alpha_{0\textbf{k}}  +  2 \epsilon_2 \hat\alpha_{0\textbf{k}2+}\hat\alpha_{0\textbf{k}2-}     
\end{eqnarray}        
with a solution, approaching:
\begin{eqnarray} \label{solution:net}
 w_{\textbf{s}\textbf{r}}=\alpha Re\left\lbrace \left(e^{i(\textbf{r}-\textbf{r}_0)\textbf{k}} +e^{i(\textbf{r}-\textbf{r}_0)\textbf{k}2-}+e^{i(\textbf{r}-\textbf{r}_0)\textbf{k}2+}\right)  \right\rbrace   
\end{eqnarray}
This means that if the strength of GC-GC connections is sufficient, then after learning grid fields of all GCs will have the same orientations and phases. It could be hypothesized that in the case of less symmetric or more noisy connectivity matrix $ \textbf{J} $ the drift of phases could be observed at scales larger than a characteristic length of connection in the GCs layer.

\section{Results}

\begin{figure}[t]
\centering
\includegraphics[width=0.8\textwidth]{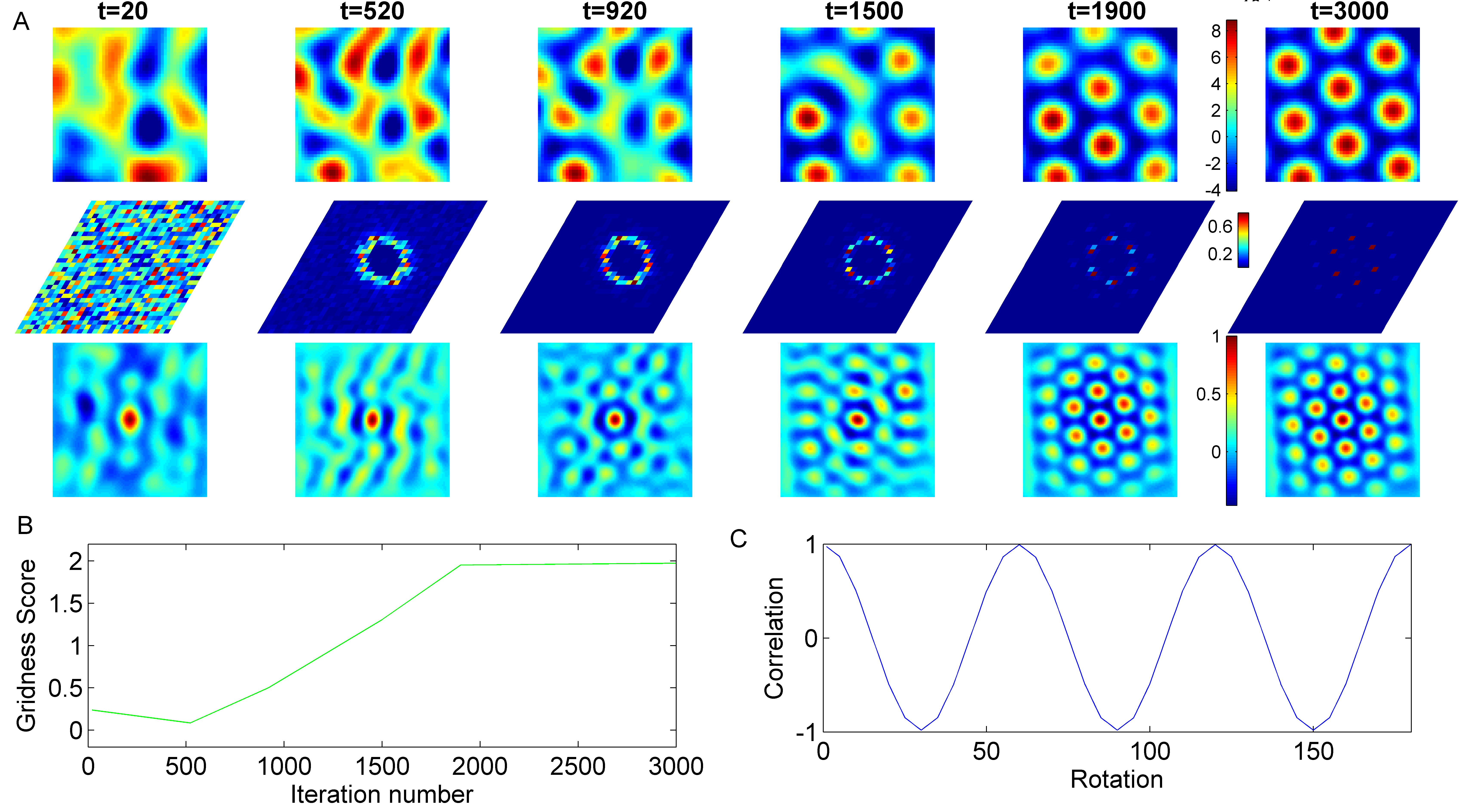}
\caption{Self-organization of grid fields in a simplified model of neuron with associative plasticity. Example of development of hexagonally symmetric fields in the model described by Eq.(\protect\ref{learning_integral}) in the twisted torus domain.
\textbf{A}. Evolution of grid fields (upper row, spatial distribution of firing activity in central part of environment is shown, height of the square is equal to the height of diamond-shaped environment) wave amplitudes (i.e. distribution of the absolute values of variables $ \alpha_{\textbf{k}} $ from Eq.(\ref{waves_of_weigths}) in wavevectors space $ \textbf{k}$ , middle row) and autocorrelation functions of grid fields (lower row). \textbf{B}. Dependence of gridness score on simulation iteration number. \textbf{C}. Pearson's correlation coefficient of autocorrelation function with its rotation transform  is shown as a function of rotation angle (\citep{Sargolini_2006, Si_Treves_2012}). }
\label{fig:1}
\end{figure}

To confirm previous analysis we have conducted a series of computational experiments. First, we have found that formation of hexagonally symmetric fields can be observed for a broad ranges of the second order terms coefficients $ \eta_{2+} $ and $ \varepsilon_2 $ in Eq.(\ref{learning_integral}). Typical results of computational experiments are shown in Fig.~\ref{fig:1}.

Linear neuron with 961 synapses was modeled. Each synapse received spatially modulated input from a PC with Gaussian-shaped firing field centered on a vertex of 31x31 hexagonal lattice embedded in diamond-shaped environment. It can be seen that during formation of grid fields (Fig.~\ref{fig:1}A, upper row) wave amplitudes (Fig.~\ref{fig:1}A, middle row) rapidly concentrate on a narrow ring, which corresponds to the critical wavelength, and then selection of single group of six waves of equal amplitudes and with wavevectors composing a hexagon occurs. Corresponding autocorrelation functions of grid fields are shown in Fig.~\ref{fig:1}A (lower row). Dependence of gridness score on simulation time step is shown in Fig.~\ref{fig:1}B. Pearson's correlation coefficient of autocorrelation function with the same function after rotation was calculated. It is shown as a function of rotation angle in Fig.~\ref{fig:1}C. Gridness score was estimated as a difference between the mean value of this correlation coefficient calculated for rotation angles  $60^0$ and $120^0$ , and its mean value for angles  $30^0$,$90^0$ and $150^0$ \citep{ Sargolini_2006, Si_Treves_2012}.

Model parameters (see Eq. (\ref{waves_evolution_eq})) used for simulations shown in Fig.~\ref{fig:1} were the following: $ \sigma=1 $, $ \varepsilon_2=50 $, $ \eta_{2+}=0 $, $\varepsilon_{1}=0  $, $ \eta_{+}=1 $, $ \eta_{-}=1 $ and $ \sigma_{\textbf{r}}=0.08L $. With these parameters maximal contribution of second order (with respect to synaptic weights) terms to the weight change rate was less than 4\% of first order terms contribution. Formation of grid fields with high gridness score (more than 1) could be observed for a broad range of values of parameter $ \varepsilon_2 $  - from 0.001 to 100. At larger $ \varepsilon_2 $  values  unimodal fields usually form instead of grid fields. As a result, those contribution of second order terms to the weight change rate for which grid field patterns formation can be observed generally do not exceed 10\%. Similar results can be observed in the case when only BCM-like second order term is nonzero ($ \eta_{2+}>0 $, $  \varepsilon_2=0 $). 

The formation of grid fields by the mechanism described above could be observed in the model Eq.(\ref{learning_integral}) for the cases when restrictions on the topology of Eq.(\ref{learning_integral}) domain used for Eq.(\ref{waves_evolution_eq}) was violated. For example, when the shape of environment was rectangular and place field centers were located on rectangular lattice, grid field with a high degree of hexagonal symmetry were formed if sizes of place fields were not too big (less than 20\% of the size of environment) and coefficients near the second order terms of Eq.(\ref{learning_integral}) were sufficiently large (they were selected close to the transition to unimodal pattern formation mode) (supplementary Fig.S1) 

\begin{figure}[t]
\centering
\includegraphics[width=0.8\textwidth]{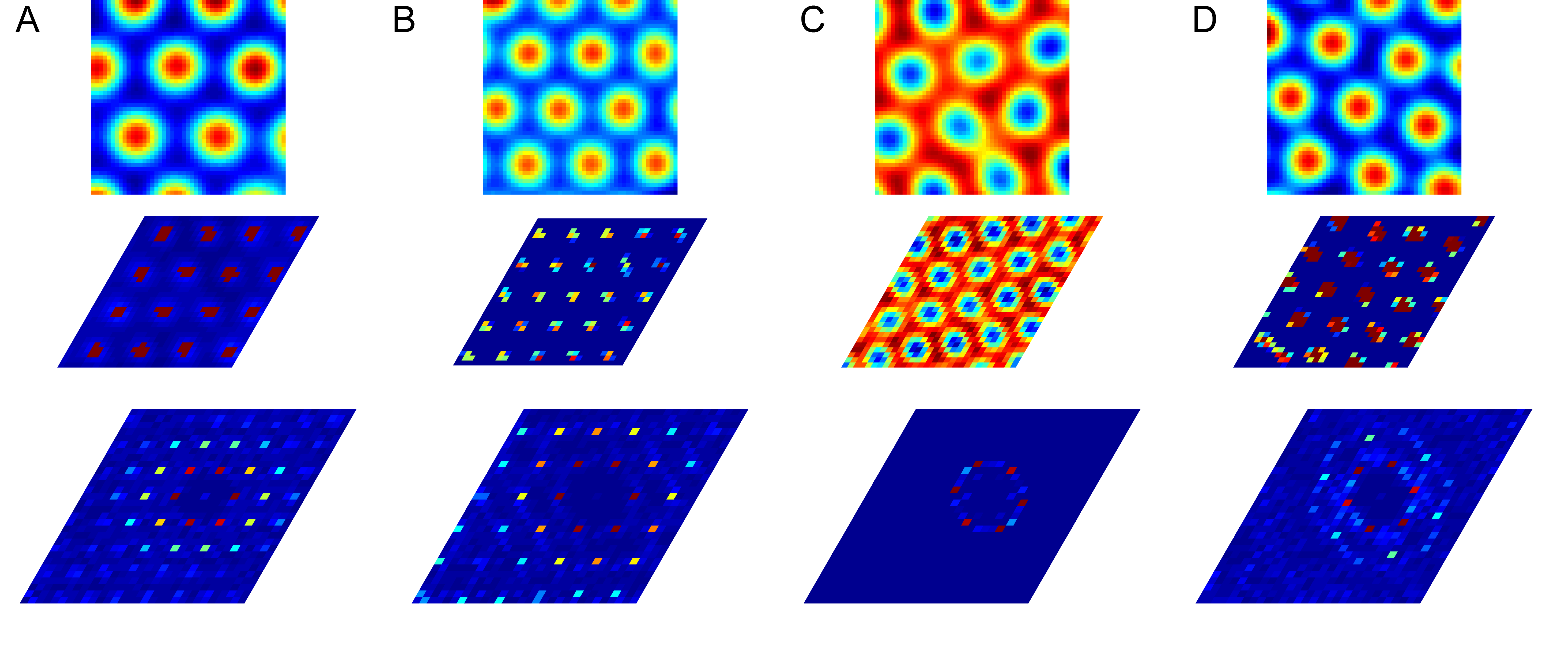}
\caption{Formation of grid fields in the models obtained from the basic model of neuron with associative plasticity by different modifications of terms describing local homeostatic plasticity  (Eq.(\ref{learning_integral})). Top row - Grid fields formed after 10000 time steps of simulation. Middle row - corresponding synaptic weights as a function of spatial location of presynaptic PCs fields. Lower row - Fouirer transforms of synaptic weights spatial distributions shown in the middle row. \textbf{A}. Imposing upper and lower bounds on the synaptic weights enable grid fields formation in the models with large second order terms. Parameters of the model (Eq.(\ref{learning_integral})) were: $ \sigma=1 $, $ \varepsilon_2=2500 $, $ \eta_{2+}=0 $, $\varepsilon_{1}=0  $, $ \eta_{+}=1 $, $ \eta_{-}=1 $ , $ \sigma_{\textbf{r}}=0.13L $. In addition, synaptic weights were limited from above by setting $ \dot{w}_{i}=0 $ for $ w_{i}>w_{max} $, $ w_{max}=0.005 $. \textbf{B}. Imposing only lower bound on synaptic weights enable grid fields formation in models without other homeostatic plasticity terms. Parameters of the model (Eq.(\ref{learning_integral})) were the same as in A except for: $ \sigma=0 $, $ \varepsilon_2=0 $, $ \eta_{2+}=0 $, $\varepsilon_{1}=0  $, $ \eta_{+}=1 $, $ \eta_{-}=1 $ , $ \sigma_{\textbf{r}}=0.13L $, $ w_{min}=-0.001 $. \textbf{C}. Models with upper limit imposed on synaptic weights learn inverted grid fields. Model parameters were the same as in A except for $ \varepsilon_2=0 $ and $w_{max}=0.005  $. \textbf{D}. In the models with both upper and lower bounds imposed on synaptic weigths, clusters of synapses are formed in which size of grid fields and size of clusters can be regulated independently. Parameters of the model were the same as in A except for: $ \varepsilon_2=0 $, $ \sigma=0 $ , $w_{min}=-0.001  $, $w_{max}=0.005  $  }
\label{fig:2}
\end{figure}

Restrictions on the maximal value of second order terms can be weakened by introducing additional plasticity rules to restrict maximal or minimal weights of individual synapses. For example, if the following parameter values are set: $ \sigma=1 $, $ \varepsilon_2=2500 $, $ \eta_{2+}=0 $, $\varepsilon_{1}=0  $, $ \eta_{+}=1 $, $ \eta_{-}=1 $ , $ \sigma_{\textbf{r}}=0.13L $  in Eq. (\ref{waves_evolution_eq}) and synaptic weights are additionaly limited from above by setting $ \dot{w}_{i}=0 $ for $ w_{i}>w_{max} $, $ w_{max}=0.005 $, then maximal ratio of second order term of Eq. (\ref{waves_evolution_eq}) to the first order terms at the end of learning is approximately equal to 1. The resulting grid fields are shown in Fig.~\ref{fig:2}A.  
It is interesting to note that introduction of limits on maximal or minimal weights into Eq.(\ref{waves_evolution_eq}) lacking second order terms ($ \varepsilon_2=0 $, $  \eta_{2+}=0 $) is sufficient to obtain the model which could learn grid-like fields. Examples of grid fields obtained for such models are shown in Fig.~\ref{fig:2}B-D. Parameter sets used in these simulations were the same as for Fig.~\ref{fig:2}A but with  $ \varepsilon_2=0 $ and $w_{min}=-0.001  $,  $ \sigma=0 $ for Fig.~\ref{fig:2}B, $ \varepsilon_2=0 $ and $w_{max}=0.005  $ for Fig.~\ref{fig:2}C and  $ \varepsilon_2=0 $, $ \sigma=0 $ , $w_{min}=-0.001  $, $w_{max}=0.005  $ for Fig.~\ref{fig:2}D. 

\begin{figure}[t]
\centering
\includegraphics[width=0.5\textwidth]{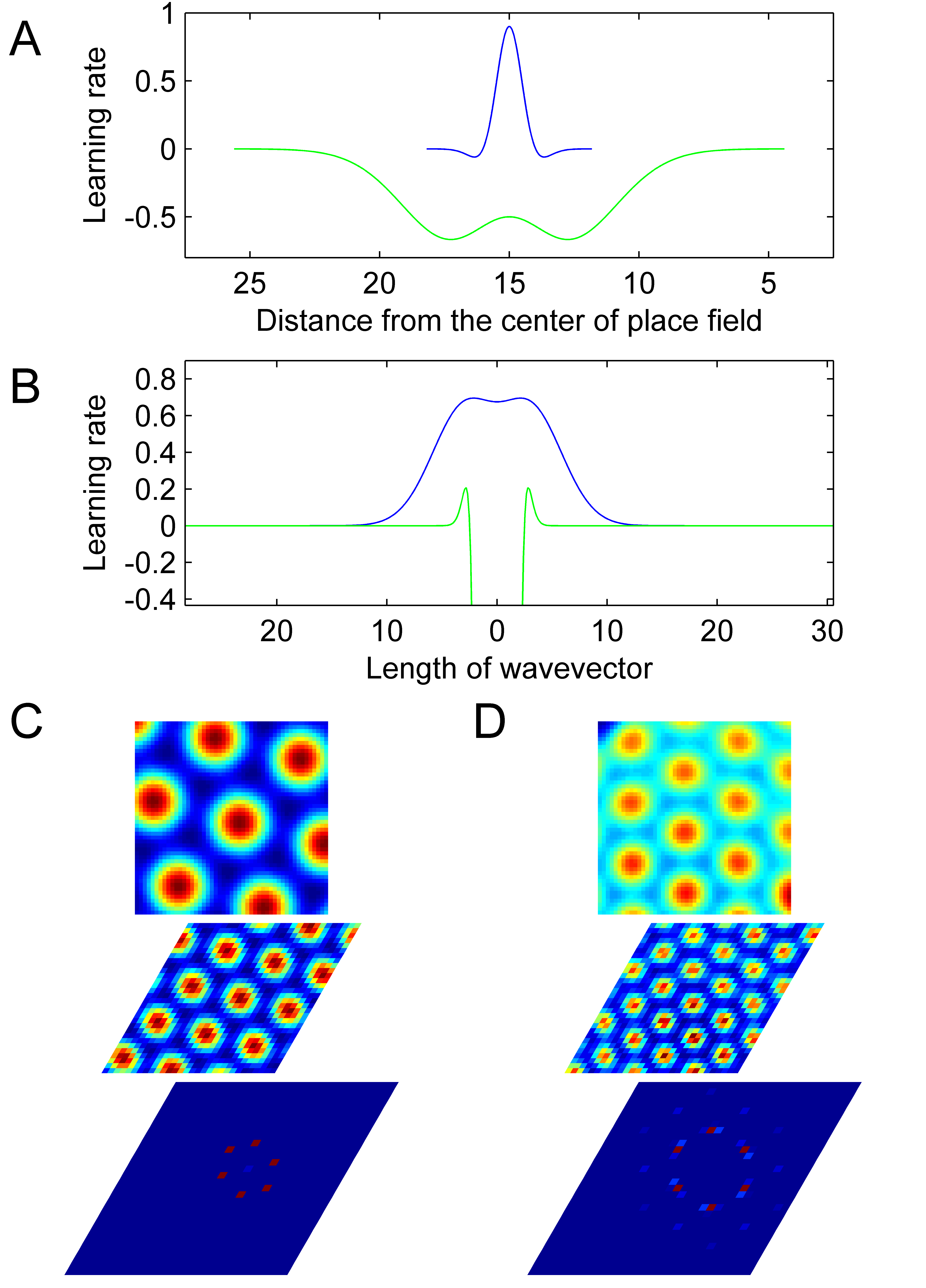}
\caption{Formation of grid fields in the models with different ratios of potentiation and depression components of associative synaptic plasticity. Ratio of depression and potentiation components was regulated by changing parameter $\eta_{-}  $ from $ \eta_{-}=0.6\eta_{+} $ (blue lines in Fig.~\ref{fig:3}A,B and Fig.~\ref{fig:3}C ) to $ \eta_{-}=2\eta_{+} $ (green lines in Fig.~\ref{fig:3}A,B and Fig.~\ref{fig:3}D)  
\textbf{A}. Dependencies of PC-GC synapse plastcity rate on the animal's distance from the center of corresponding place fields at a fixed postsynaptic firing rate. It can be seen that in the case of small $ \eta_{-} $ the maximal value of negative plasticity rate (which could be considered as those resulting from LTD) is less than 7\% of maximal positive plasticity rate. In the case of large $ \eta_{-} $ all plastic changes could be considered as resulting from LTD.
\textbf{B}. Corresponding dependencies of wave amplitude change rates on the length of its wavevectors. \textbf{C},\textbf{D}.Grid fieds (top row), synaptic weights as a functions of spatial position of the centers of place fields (middle row) and wave amplitudes as a function of wavectors (lower row).  }
\label{fig:3}
\end{figure}

On the contrary to the low sensitivity of grid field formation process to the second order terms coefficients, the ratio of first order terms coefficients could vary approximately within the limits of $ \eta_{-}=0.6\eta_{+} $ and $ \eta_{-}=2\eta_{+}  $. But these limits correspond to physiologically different situations - Hebbian plasticity with large LTP component and very small LTD component for small $  \eta_{-} $ values and pure LTD with nonlinear dependence on presynaptic activity for large $  \eta_{-} $ values. Examples of plasticity rate curves and grid fields obtained for different ratios of $ \eta_{-}$ to $\eta_{+}  $ are shown in Fig.~\ref{fig:3}. Here in Fig.~\ref{fig:3}A dependencies of plasticity rate of a PC-GC synapse on the distance of animal's location from the center of corresponding place fields at fixed postsynaptic firing rate are shown for $ \eta_{-}=0.6\eta_{+} $ and $ \eta_{-}=2\eta_{+}  $ by blue and green lines, respectively. It can be seen that in the case of small $ \eta_{-} $ the maximal value of negative plasticity rate (which could be considered  as a result of LTD) is less than 7\% of maximal positive plasticity rate. In the case of large $ \eta_{-} $ all plasticity changes could be considered as a result of LTD. Corresponding dependencies of wave amplitude change rates on the length of its wavevectors are shown in Fig.~\ref{fig:3}B. Grid fields (top row), synaptic weigths as a function of spatial position of the centers of place fields (middle row) and wave amplitudes as a function of wavectors (lower row) are shown for the case $ \eta_{-}=0.6\eta_{+} $ in Fig.~\ref{fig:3}C and for the case $ \eta_{-}=2\eta_{+} $ in Fig.~\ref{fig:3}D. 

\begin{figure}[t]
\centering
\includegraphics[width=0.8\textwidth]{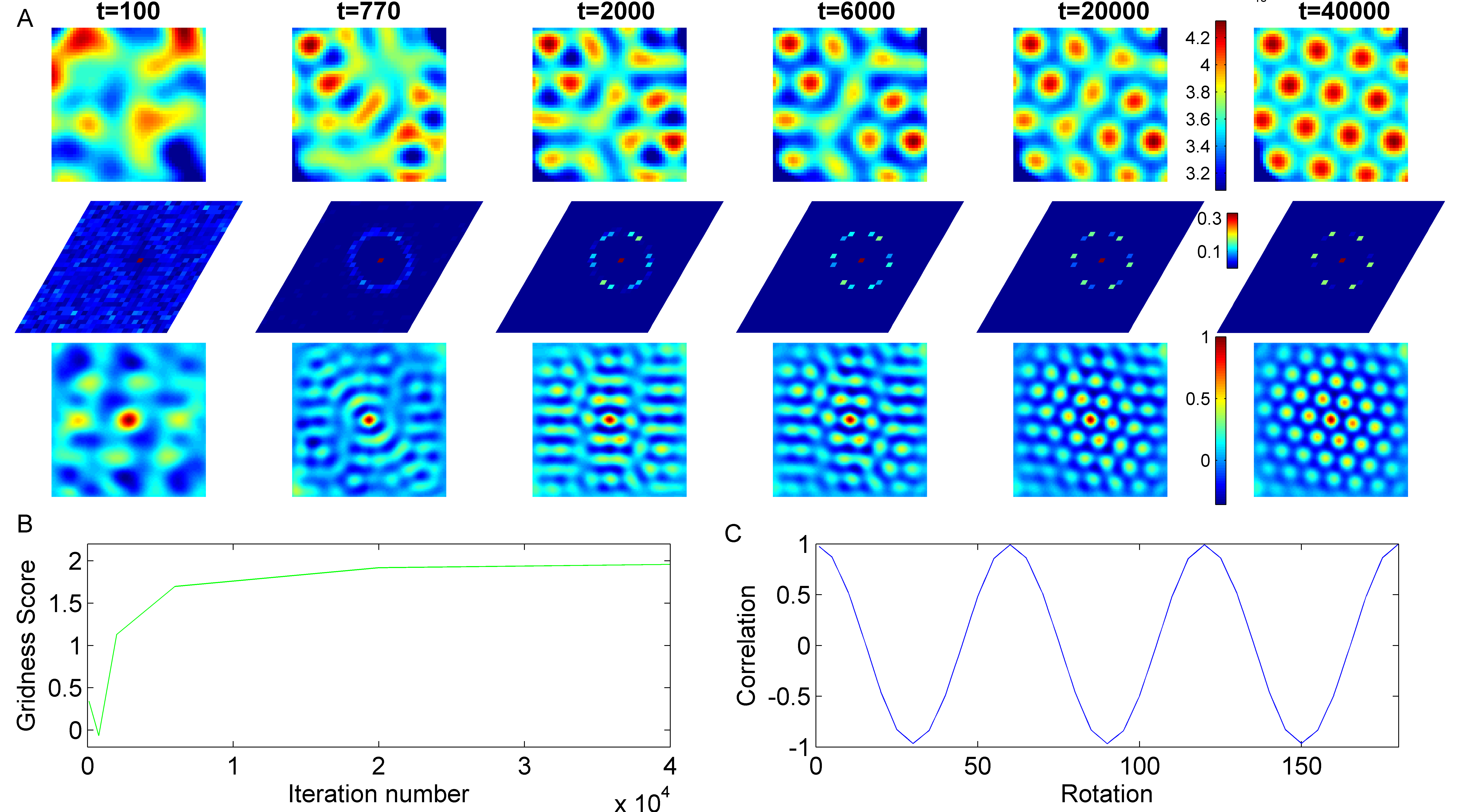}
\caption{Learning of grid fields in the model with sublinear LTD and BCM-like LTP terms without additional homeostatic plasticity restrictions on individual synaptic weigths. Panels are the same as in Fig.~\ref{fig:1}.}
\label{fig:4}
\end{figure}

It is possible to modify the model in such a way that formation of grid fields becomes possible due to interaction of LTD and BCM-like LTP terms without additional homeostatic plasticity restrictions on individual synaptic weights. This can be achieved, for example, if dependence of LTD rate on presynaptic activity is sublinear. Example of learning in such model is shown in Fig.~\ref{fig:4} . 
Learning equation was:
\begin{eqnarray} \label{BCM_lr}
  \dot{w}_{i} =  -\eta_- \textbf{R}_{0i}\textbf{w}+ \eta_{2+}\textbf{w}^T\textbf{Q}_i\textbf{w}   -\sigma \sum_{k}{w_{k}^2}w_{i}+\varepsilon_{1}
\end{eqnarray}
with parameters:  $ \sigma=100 $, $ \eta_{2+}=5000 $, $\varepsilon_{1}=5  $,  $ \eta_{-}=100 $ , $ \sigma_{\textbf{r}}=0.13L $ . First term of this equation was obtained by integrating over time the expression describing nonlinear LTD:
\begin{eqnarray}  
\sum_{t}{\eta_-y_{t}\sqrt{x_{it}}}\approx-\eta_- \textbf{R}_{0i}\textbf{w}
\end{eqnarray}
where 
\begin{eqnarray}
{R_{0ij}= \int F_i(\textbf{r})^{\frac{1}{2}}F_j(\textbf{r})\textbf{r}^D = F_0\exp\left( -\frac{1}{6\sigma_r^2}\left\| \textbf{r}_i-\textbf{r}_j \right\|^2\right)}   
\end{eqnarray}
It should be noted that learning of grid fields in this model was possible only for narrow range of parameters. For example, the range of parameter $ \varepsilon_{1} $ values, for which the formation of grid fields was observed was between 3 and 5.

To confirm results obtained for simplified model described by Eq.(\ref{learning_integral}) in more realistic conditions we have modeled grid fields formation as a result of offline learning after environment exploration. In these simulations we assume that after exploration of diamond-shaped environment by an animal, episodic memory corresponding to the visited places could be spontaneously reactivated in such a way that relative activities of place cells during reactivation are similar to those at time of navigation. The physiologiacal basis for such reactivation could be for example a sharp wave activity observed in hippocampus during sleep and quiet wakefulness periods \cite{Csicsvari_2014}.

We have observed that for sufficiently long path of movement, the process of spatially modulated fields formation was qualitatively similar to those described above for simplified model.
For these simulations we directly solve equation obtained by simplification of Eqs.(\ref{activation})-(\ref{asslearning_gen}). After setting  $  \zeta(\textbf{w}_t)=-\sigma \sum_{k}{w_{kt}^2}$ and  $ P(w_{it},x_{it})=BH(x_{it}-\theta)w_{it}^2  $ in Eqs.(\ref{hom_plasticity:1})-(\ref{hom_plasticity:2}) we obtain:
\begin{eqnarray} \label{offline_learning}
 \dot{w}_{it}= -\eta_- x_{it}y_{t}+ \eta_{2+}x_{it}y_{t}^2 + \eta_+ x_{it}^2y_{t}   -\sigma \sum_{k}{w_{kt}^2}w_{it}+Bw_{it}^2H\left(x_{it}-\varTheta \right) 
\end{eqnarray}
where $ H $ is the Heaviside function.
Offline learning was modeled by calculating synaptic weight changes at each iteration of the algorithm independetly for 5000 randomly selected places on the path of movement, and then updating synaptic weights using the sum of these weight changes.
The set of parameter values was $ \sigma=1 $, $ B=1000 $, $ \eta_{2+}=0 $,  $ \eta_{+}=1 $, $ \eta_{-}=3 $, $ \varTheta=0 $.
Place fields were modeled with Gaussian functions $  F_i(\textbf{r})\approx  F\exp(-\frac{1}{2\sigma_r^2}\left\|\textbf{r}-\textbf{r}_i\right\|^2  ) $ with $ F=20 $. Place field size $ \sigma_r $ was varied from $ 0.1L $ to $ 0.3L $, where $ L=1 $ is size of diamond-shaped environment. Centers of place fields were initially selected at vertexes of 31x31 hexagonal lattice which was embedded in environment. Then all place field centers were perturbed by addition of a random vector selected from Gaussian distribution with $ \sigma=0.5\varDelta L $ ($ \varDelta L  $ - lattice spacing).

Path of movement was obtained by the integration of the movement equations:
\begin{align} \label{movement_eq}
\textbf{r}_{t+\Delta t}=\textbf{r}_t+\textbf{v}_t \Delta t\\
\textbf{v}_{t+\Delta t}=0.99\textbf{v}_t+0.01\delta \textbf{v}_t ,\quad \delta \textbf{v}_t\sim\mathcal{N}\left(0,1 \right) \nonumber
\end{align}
At the borders of environment direction of motion was selected randomly and the speed $ \left\| \textbf{v}_t \right\|  $ was set to 0.1 of its value just before collision with a border. Time step used for the integration of movement equations was $ \Delta t=0.1 $. Average speed of motion along the path obtained after the solution of Eq.(\ref{movement_eq}) was $ \left\langle \left\| \textbf{v}\right\|  \right\rangle =0.06 $ and the coefficient of variation of speed was $ CV=0.64 $.   

\begin{figure}[t]
\includegraphics[width=0.98\textwidth]{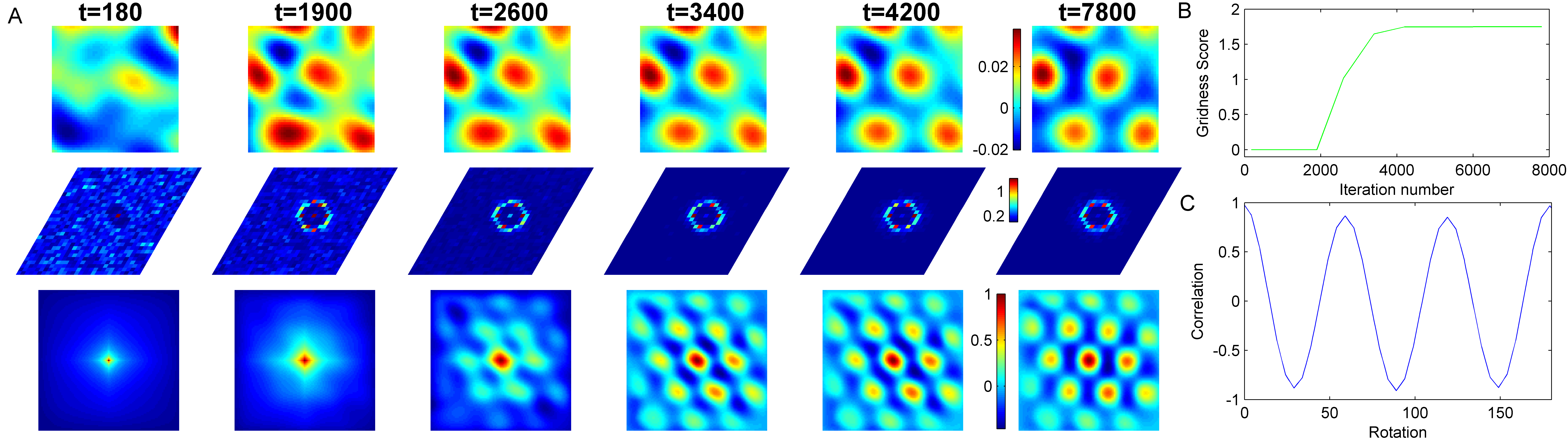}
\caption{Simulation of grid fields offline learning in the basic model of EC neuron resulting from memory reactivation after animal's exploration of environment.
Random path of the length $ S=172L $ in a diamond-shaped labirinth of size $L=1  $ was simulated. Then patterns of PCs activity observed during navigation were reactivated in random order during the period of offline learning. Panels are the same as in Fig.~\ref{fig:1}. }
\label{fig:5}
\end{figure}

Typical result of simulations is shown in Fig.\ref{fig:5}. 
The total length of traveling path in a diamond-shaped environment of size L=1 was S=172. It can be seen that similarly to the case of model described by Eq.(\ref{learning_integral}), formation of grid fields occurred in two stages: initially, a ring of waves with critical wavelength grew, then selection of six waves whose wavectors composed a hexagon took place. The difference with respect to simplified model is that not all waves out of the hexagon of waves with maximal amplitudes dissapear completely at the end of learning and that the amplitudes of waves composing the hexagon are not equal. As a result, the gridness score of observed fields vary in a broad range and highly depends on grid size and path length.

Learning of smaller size grid fields generally requires significantly longer paths. We observed that this requirement can be significantly weakened by introducing restrictions on a minimal weight of each synapse. We observed that in this modified version of the model fast self-organization of grid fields could be achieved during online learning.  

To demonstrate this we have conducted a series of computational experiments. In these experiments the shape of environment was a square with a side L=1, place fields had nonperiodic Gaussian shape, and positions of their centers were selected randomly within the environment. Online learning was achieved by solving Eqs.(\ref{activation})-(\ref{asslearning_gen}) with global homeostatic plasticity switched off and local homeostatic plasticity in the form of lower bound on synaptic weights (i.e. equation $ P(w_{it},x_{it})=-BH(w_{min}-w_{it}) $ was used instead of Eqs.(\ref{hom_plasticity:2})). Online learning equation was:
\begin{eqnarray} \label{online_learning}
 \dot{w}_{it}=\kappa_t \left( -\eta_- x_{it}y_{t}+ \eta_+ x_{it}^2y_{t}  -BH\left(w_{min}-w_{it} \right)\right)  
\end{eqnarray}
Parameter values were:  $ \eta_{+}=1 $, $ \eta_{-}=1.125 $, $ w_{min}=-0.2 $, $ B=1000 $. 
Decreasing learning rate was important for the fast formation of grid fields during online learning and at the same time for achievment of a high gridness score at the end of learning. Parameter $ \kappa_t $ describes learning rate changes with time. It was selected in such a way that root mean square change of synaptic weights from a time moment $ t$ to $ t+\Delta t $ was equal to  $ \sqrt{\left\langle \Delta w_t^2\right\rangle} =\frac{1.6\cdot 10^{-3}}{1+0.01t}+1.6\cdot10^{-4}$. Time step used for the integration of online learning equation was $ \Delta t=0.1 $. 
Place fields were modeled as: $  F_i(\textbf{r})\approx  F\exp(-\frac{1}{2\sigma_r^2}\left\|\textbf{r}-\textbf{r}_i\right\|^2  ) $ with $ F=1 $. 

\begin{figure}[b]
\centering
\includegraphics[width=0.98\textwidth]{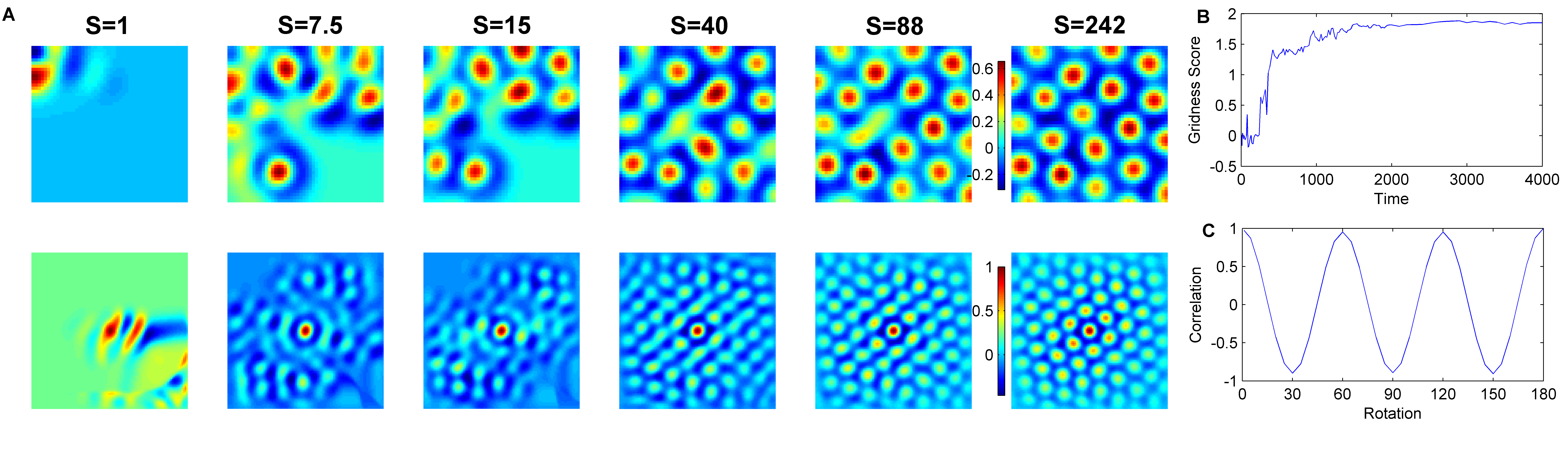}
\caption{Simulation of grid fields online learning in the  model of EC neuron with bounded synaptic weights.
Random path of the length $ S=242L $ in a square labirinth of size $L=1  $ was simulated. The GC with 1000 synapses from PCs with place fields centers randomly distribured in environment was modeled. Panels are the same as in Fig.~\ref{fig:5}.   }
\label{fig:6}
\end{figure}

Example of grid fields formation during online learning experiment is shown in Fig.~\ref{fig:6}.

\begin{figure}[t]
\centering
\includegraphics[width=1\textwidth]{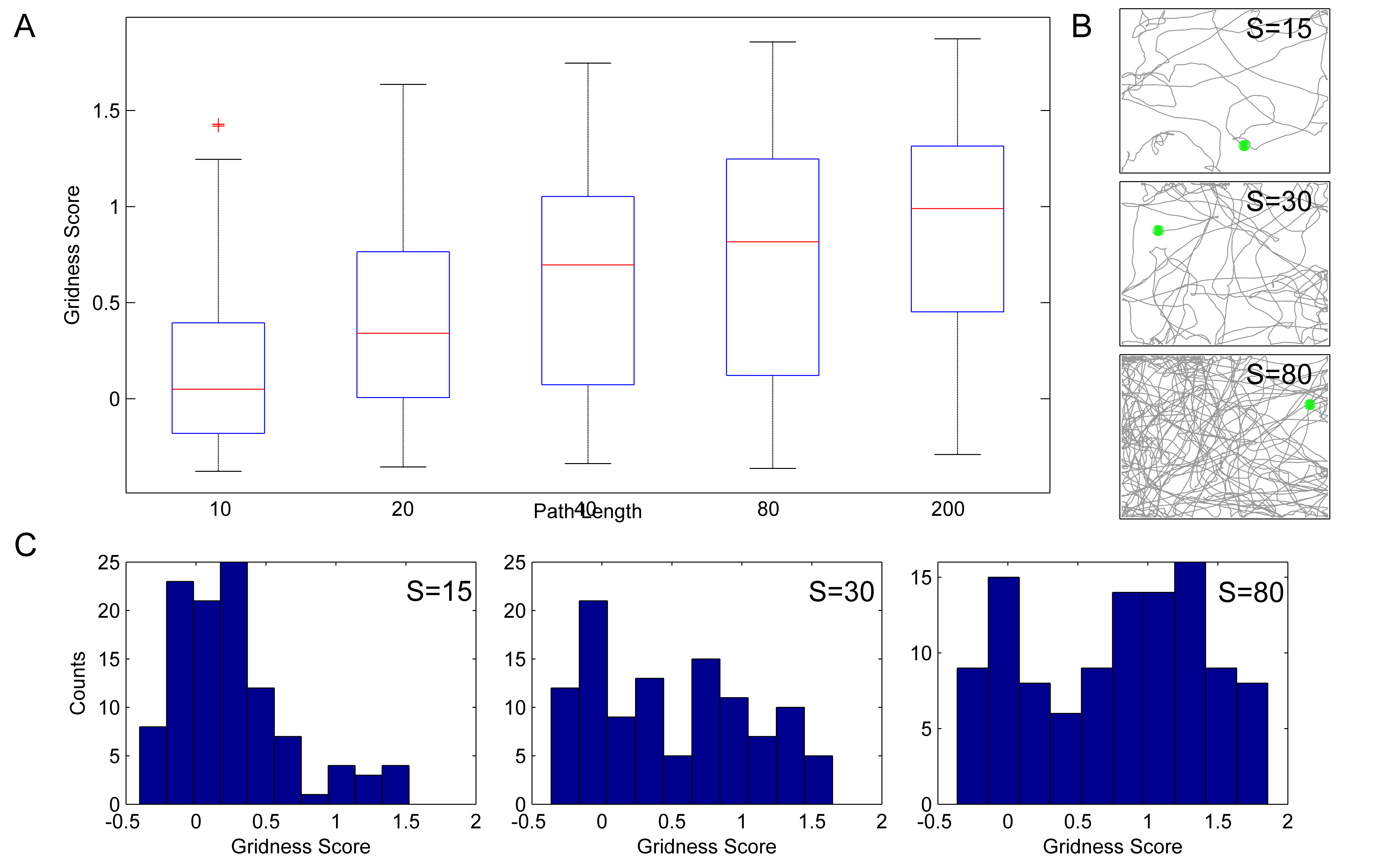}
\caption{Statistics of the quality of grid fieds observed in online learning simulations. 
\textbf{A}. Statistical box plots represent statistics of gridness scores for the grid fields obtained in a series of N=100 online learning simulations for a model of EC neuron with bounded synaptic weights (Eq.(\ref{online_learning})). Each box corresponds to the animal's movement paths of a different lengths embedded in the square environment of the size $ L=1 $. On each box, the central red line coresponds to the median gridness score, the edges of the box are the 25th and 75th percentiles, the whiskers extend to the most extreme data points not considered outliers, and outlier is indicated by the red cross.
\textbf{B}. Examples of paths that were used in experiments described in A (represented by gray line, green circle - current position of the animal), corresponding to the mean path lengths 15 (top), 30 (middle) and 80 (down). \textbf{C} Histograms represents distributions of gridness scores obtained in online learning experiments (N=100) corresponding to the mean path lengths 15 (left), 30 (middle) and 80 (right). }
\label{fig:7}
\end{figure}

We found that in these experiments probability to obtain fields of size $ >0.2L $ with gridness score higher than $ 0.5 $ at the end of online learning was high (in 69 cases of 100 experiments, gridness score in these 69 experiments was  $Mean\pm SD=1.19\pm0.3$). Most part of gridness score growth (67\%) in simulations with high score ($ >0.5 $) at the end of learning was achieved during first 10 minutes of motion (with average speed $ v=0.06L/s $).
The dependence of gridness score on the traveling path is shown in Fig. ~\ref{fig:7}A. 

Here, statistical boxes correspond to the results of simulations of online learning for N=100 different paths for 5 different path lengths. Examples of movement trajectories for path lengthes S=15 (upper panel),30 (middle panel) and 80 (lower panel) are shown in Fig.~\ref{fig:7}B. Distributions of gridness scores obtained in online learning experiments corresponding to the mean path lengths 15 (left), 30 (middle) and 80 (right) are shown in Fig.~\ref{fig:7}C.  
It can be seen from these experiments that large-sized fields ($ \sigma_r>0.2L $, were $\sigma_r $- size of grid field) with relatively high positive gridness score ($>0.5$) could be formed with high probability (19 of 100) after relatively short experience of navigation in a new environment (mean path length - $ S=15L $, mean time of motion with average speed $ v=0.06L $ - $ T=240$ ) (Fig. ~\ref{fig:7}B, upper trace). 
Additional experiments demonstrated that formation of grid fields with relatively high gridness score was observed approximately for the same range of LTD/LTP ratios as those for which grid fields formation was possible in the basic model (Eq.\ref{learning_integral})in the twisted torus domain. Sizes of grid fields formed varied from 11 for LTD/LTP=2.8 to 23 for LTD/LTP=0.8 for the rectangular environment of size L=50 (Fig.S2 and Fig.S3).

\section{Discussion}
In this work we have described formation of grid-like spatially selective sensory fields in the group of models of EC neuron having linear activation function and recieving spatially modulated input from place cells or from neurons of different sensory systems.

We have shown that learning of hexagonaly symmetric fields in these models is a result of complex action of several simple biologicaly motivated synaptic plasticity rules. These rules include rules for associative synaptic plasticity similar to those widely used for description of linear LTD and LTP rule with supralinear dependence on presynaptic activity, or a combination of BCM-like rule for LTP with sublinear LTD rule. We observed that even solely LTD rule with nonlinear dependence of LTD rate on presynaptic activity, or a rule with very small LTD in comparison with LTP component, is sufficient for robust grid-like fields formation.

From the perspective of pattern formation theory, the role in learning of the rules for associative synaptic plasticity with linear dependence on synaptic weigths is selection of patterns composed of a group of waves with similar wavelengthes. This critical wavelength determines the size of grid fields.\\
Additional plasticity rules with nonlinear dependence on synaptic weights are required for selection of subgroup of six waves, whose wavevectors compose a hexagon, from the group of waves with critical wavelength. These rules can be a consequence of restrictions imposed on maximal or minimal synaptic weigths, of a small supralinearity of the dependence of synaptic plasticity rate on the synaptic weight, or nonlinearity of a neuron activation rule. BCM-like LTP term nonlinearly depends on synaptic weigths and as a result it can, together with sublinear LTD term, play a role in both selection of waves with critical wavelength and consequent formation of hexagonally symmetric pattern  (Eq. \ref{BCM_lr}). 

In addition to these results we have shown that within the basic model (Eq.\ref{offline_learning} and Eq.\ref{learning_integral}) grid fields formation could be observed in experiments with a simulation of random movement in the environment, but learning in this model was slow because of restriction of the maximal values of second order terms of learning equations. 

On the other hand, the grid fields self-organisation model composed of LTD-like and LTP-like linear, with respect to synaptic weights, terms, and additional nonlinear local homeostatic plasticity term, which restrict minimal synaptic weight (Eq.(\ref{online_learning})) , are capable of fast online learning of grid fields after relatively short expirience of navigation in a novel environment. ). The quality of grid fields was not lower than in offline learning experiments (for the same path length). The main difference between offline and online learning is that the latter requires selection of an appropriate schedule for learning rate changes – at learning rates for which fast self-organization could be observed grid fields do not stabilize with a time. For online learning experiments described in the manuscript learning rate with a hyperbolic dependence on time was used (with ratio 10:1 between initial and final values). We found that main part of online learning could occur rapidly – within 10-20 minutes of navigation in 1m labyrinth with average speed  6cm/s (for 20cm fields ). Learning rate was the only parameter which should be additionally tuned for online learning. Nevertheless we assume that offline learning scenarios could be considered as a model of interaction of hippocampus with entorhinal cortex during hippocampal sharp-wave activity \cite{Csicsvari_2014}.  

The model described in this work differs from other published models of grid fields formation that are based on synaptic plasticity in PC-GC synapses. First, it does not require additional firing rate adaptation process as it is supposed to be important for grid fields learning in the model of Korpff and Treves {\citep{Kropff_Treves_2008,Si_Treves_2012}}. As a result, learning in the proposed model is not sensitive to the inhomogeneity and anisotropy of the animal's motion velocity distribution. In addition, the model allows learning on the basis of random reactivation of episodic memory of navigational information without requirment for recall of continuos fragments of movement trajectory.

The speed of the grid field formation in modications of the proposed model, in which the lower boundary on synaptic weights is added (Eq.(\ref{online_learning})), is similar to those observed for the model described in the work of Castro and Aguiar {\citep{Castro_2014}}, but learning rules used in our model have more general form and are better biologically motivated.

Recently proposed model of Widloski and Fiete \cite{Widloski_2014} significantly differs from the model proposed in this work. First, it assumes that EC neurons have unimodal place fields pre-existing before learning begins, and only the lateral EC synapses are modified during learning. Second, STDP rules are used in their model and different rules are required for different subpopulations of synapses. Mexican hat type lateral synaptic connections are formed as a result of learning and grid-like activity in their model is a result of a multi-bump attractor existing in the model after learning. It is interesting to note that frequency-dependent synaptic plasticity rule used in our work could be used instead of STDP rule within Widloski and Fiete \cite{Widloski_2014} model for learning of Mexican hat connectivity, and, if augmented with additional adaptation mechanism - for bump control learning.

Some of the models (Eq.\ref{learning_integral}) described in this work allowed detailed mathematical analysis. This property could be useful for the analysis of possible self-organization mechanism of modular structure of grid cell networks observed in experiments. These structures assume that several specific patterns of GCs network activity, including alignment of grid field orientations, should arise in a network of GCs. Using theoretical analysis we have demonstrated that models of GCs network composed of cells with plastic PC-GC connections and fixed weak GC-GC connections could align orientations of their grid fields. This result is similar to those described in the works {\citep{Kropff_Treves_2008, Si_Treves_2012}}.

From the experimental observations it is known that distribution of sizes of grid fields observed in MEC has discrete spectrum \citep{Stensola_2012}. This property could have a simle explanation on the basis of the theory proposed in our work. It can be seen that the spatial distribution of synaptic weigths, which forms in the model with lower-bounded synaptic weights (Eq.\ref{online_learning}), has a spectrum consisting of peaks which form the hexagonal lattice (Fig.2B, middle panel). To reproduce discretness of grid field sizes, we propose a network consisting of a layer of neurons in EC (disconnected or weakly connected with each other) recieving strong synaptic input from a module of GCs with grid fields formed on the basis of model described by Eq.(\ref{online_learning}). It could be predicted that if synapses in this network have learning rule similar to PC-GC synapses rule, then GCs with a smaller grid fields sizes will be formed (with the ratio  $ sqrt(3):1 $ or higher between sizes of fields).     

In conclusion we suggest that learning on the basis of simple and biologically plausible associative synaptic plasticity rules could result in formation of grid-like fields in EC which receive synaptic inputs from hippocampal place cells and spatially modulated sensory inputs from different sensory systems. It is likely that such learning contributes to the formation of grid fields in early development and to the maintainence of normal grid cells activity patterns in familiar environments.

\section*{Conclusions}
1. Learning of hexagonaly symmetric fields demonstrated in the model of EC neuron with linear activation function. The neuron recieved spatially modulated input from place cells, or cells from sensory systems. Learning rules used in the model have a general form and are biologically motivated.
  
2. From the perspective of pattern formation theory, associative synaptic plasticity, linear with respect to synaptic weigths, is required for the selection of patterns composed of waves with similar critcal wavelengthes. This critical wavelength determines the size of grid fields. Additional plasticity, which depends nonlinearly on synaptic weights, is required to make a selection of subgroup of six waves, whose wavevectors compose a hexagon, from the group of waves with critical wavelength.
 
3. The model containing LTD-like and LTP-like linear, with respect to synaptic weights, terms, and additional nonlinear local homeostatic plasticity term, demonstarte fast online learning of grid fields in a novel environment even after a short navigational expirience. 
 
4. Aligment of grid field orientations in grid cell networks and discretness of grid fields sizes observed in experiments could be explained within the proposed model.

\section*{Acknowledgments}
The author would like to thank Alexei Samsonovich for the suggestion of the idea of principal component analysis of place cells activity patterns as a possible mechanism of grid fields formation, that lay down in the basis of this work.
This work was supported by the Ukrainian government.

\section*{References}

\bibliography{mybibfile}

\end{document}